\documentclass[twocolumn,showpacs,preprintnumbers,amsmath,amssymb]{revtex4}
\usepackage{graphicx}
\usepackage{dcolumn}
\usepackage{bm}

\begin{document}
\title{Cumulation of High-current Electron Beams: Theory and Experiment}

\author{S.V.~Anishchenko}
\email{sanishchenko@inp.bsu.by}
\affiliation{Research Institute for Nuclear Problems\\
Bobruiskaya str., 11, 220030, Minsk, Belarus.}%

\author{V.G.~Baryshevsky}
 \email{bar@inp.bsu.by}
 \affiliation{Research Institute for Nuclear Problems\\
Bobruiskaya str., 11, 220030, Minsk, Belarus.}%

\author{N.A.~Belous}
 \affiliation{Research Institute for Nuclear Problems\\
Bobruiskaya str., 11, 220030, Minsk, Belarus.}%

\author{A.A.~Gurinovich}
\email{gur@inp.bsu.by}
\affiliation{Research Institute for Nuclear Problems\\
Bobruiskaya str., 11, 220030, Minsk, Belarus.}%

\author{E.A.~Gurinovich}
  \affiliation{Research Institute for Nuclear Problems\\
Bobruiskaya str., 11, 220030, Minsk, Belarus.}%

\author{E.A.~Gurnevich}
 \email{genichgurn@gmail.com}
 \affiliation{Research Institute for Nuclear Problems\\
Bobruiskaya str., 11, 220030, Minsk, Belarus.}%

\author{P.V.~Molchanov}
\email{molchanov@inp.bsu.by}
  \affiliation{Research Institute for Nuclear Problems\\
Bobruiskaya str., 11, 220030, Minsk, Belarus.}%

\begin{abstract}
A drastic cumulation of current density caused by electrostatic
repulsion in relativistic vacuum diodes with ring-type cathodes is described theoretically and confirmed
experimentally.  The distinctive feature of the suggested
cumulation mechanism over the conventional one, which relies on
focusing a high-current beam by its own magnetic field, is a very
low energy spread of electrons  in the region of maximal current
density that stems from a laminar flow profile of the
charged-particle beam.
\end{abstract}

\pacs{84.70.+p, 52.59.Mv}

\maketitle

\section{Introduction}

The pioneer research into high-current electron beams dates back
to the 30ies of the past century  \cite{Bennet1934}. For the lack
of  equipment and tools affording  the generation of high-power
 charged-particle beams under terrestrial conditions, the
 researchers mainly focused their attention  on  theoretical consideration of
 astrophysical problems \cite{Alfven1939}.

The first high-current electron beams with the power from several
of gigawatts to several of terawatts
\cite{Graybill1967,Link1967,Charbonnier1967,Graybill1971,Shipman1971}
 obtained three decades later made a revolution in the
cumulation research. This became possible primarily through two
remarkable achievements in experimental physics: First, Dyke and
colleagues experimentally obtained current densities as high as
$10^8$ amperes/cm$^2$ from the microprotrusions of the metal
cathode placed in a strong electric field; second, the dielectric
breakdown data reported by J. Martin and colleagues
\cite{Graybill1971,Martin1992} provided the potential for
developing high-voltage pulse generators.

Self-focusing of high-current electron beams with their own magnetic
fields \cite{Morrov1971,Bradley1972} provided the charged-particle
beam intensities as high as $\sim 1$\,TW/cm$^2$, thus enabling the
laboratory investigation of the extreme state of matter. The
expectation was that by cumulation of high-current beams, the
deuterium-tritium targets would be compressed and heated  to
ignition so as to initiate thermonuclear reactions and thus
accomplish pellet fusion
\cite{Yonas1974,Rudakov1974}.

Though the initially set goal of developing a pellet
fusion was not achieved, high-current electron beams found
successful applications in other fields of physics \cite{Kolb1975,
Rudakov1990, Mesyats2004}. They are used for research  in
radiation physics  ~\cite{Martin1969},  generation of  high-power
microwaves \cite{Kovalev1973, Carmel1974},  collective
acceleration of ions ~\cite{Rander1970, Dubinov2002}, and pumping
gas lasers ~\cite{Basov1986}. Nonlinear phenomena originating from
the high-current-beam interaction with self- and  external
electromagnetic fields figure prominently in all these processes.

This paper considers one of such phenomena, which is, in fact, an
alternative mechanism of high-power electron beam cumulation. This
mechanism occurs in relativistic vacuum diodes with a ring-type cathode. Even though this
phenomenon has been experimentally observed for years, it still
lacks a consistent explanation. Our main task here is
to provide a theoretical description of  this
cumulation mechanism and the experimental verification thereof. We
will show in the subsequent sections that the explosive electron
emission changing  the emitting surface of a high-current diode is
paramount for the cumulation process.

This paper is arranged as follows: We shall first give quite a
detailed description of the phenomenon of explosive electron
emission; then we shall describe the cumulation mechanism of the
electron beam that was revealed during modeling the relativistic
vacuum diode operation  with the self-developed computer code (see
\cite{AnishchenkoGurinovich2014}, the underlying computer
algorithm is given in the Appendix).  In conclusion, we shall
report the experimental results that confirm the described
cumulation mechanism for high-current electron beams.

\section{Electron emission}
High-current electron beams are generated  in relativistic vacuum
diodes, composed of  a cathode and an anode, through explosive
electron emission (see fig.1). The mechanism of the explosive
electron emission is as follows
\cite{Mesyats1971,Mesyats1975,Mesyats2005}: once the voltage is
applied across the electrodes of the relativistic diode, the
field-emission current is emitted from the cathode surface
\cite{Wood1897, Fowler1928, Millikan1929, Murphy1956}, which  is
the electrons tunneling from metal into a vacuum under the
influence of the electric field. The electrons moving in the metal
heat the cathode surface. The microscopic electric field near the
cathode is nonuniform because of the surface defects of
the  conductor. Particularly, the field near the
microprotrusion tips is sufficiently greater than the average one,
which causes rapid heating of the tips that explodes as the
specific energy density rises to $\sim 10^4$\,J/g. 

When the
macrofield becomes as large as $1$\, MV/cm, the time delay $t_d$
of the explosion is as small as 1 ns. Each microexplosion is
accompanied by thermionic emission from the surface of the cathode
flare --- conducting plasma expanding at a speed $v\sim 10^4$\,m/s
 (see Fig.2).

\begin{figure}[ht]
\begin{center}
\resizebox{85mm}{!}{\includegraphics{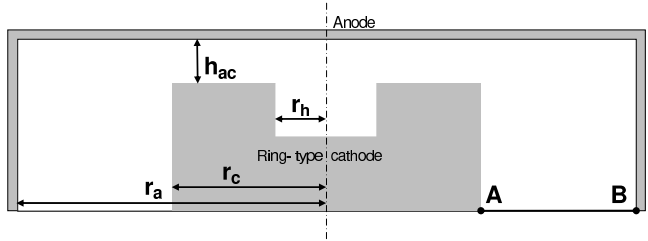}}\\
\end{center}
\caption{Relativistic vacuum diode with a ring-type cathode} \label{Fig.1}
\end{figure}

\begin{figure}[ht]
\begin{center}
\resizebox{85mm}{!}{\includegraphics{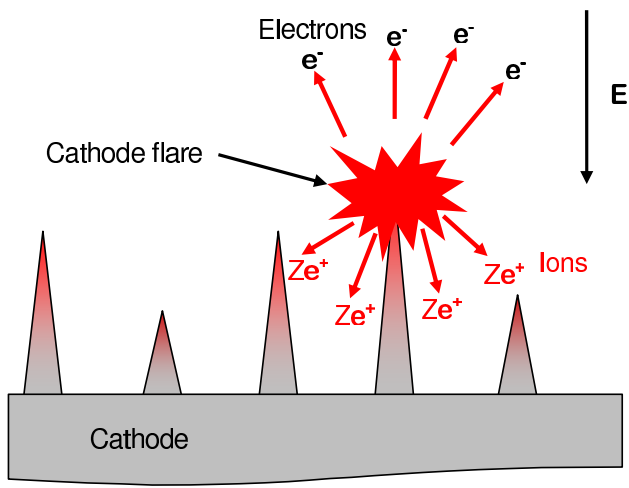}}\\
\end{center}
\caption{Explosive electron emission} \label{Fig.2}
\end{figure}

It is noteworthy that $t_d$ is particular sensitive to the
condition of the cathode surface. It was shown
in \cite{Farrall1975} that dielectric  inclusions on the
cathode surface result in the excessive increase of the
field-emission current.
B.M.Cox and W.T.Williams ~\cite{Cox1977} reported experimentally
observed high electric field strengths near dielectric inclusions
on the cathode surface with the local field strength in the
vicinity of the inclusions being several hundreds times greater
than the average field strength in the cathode-anode gap. The
dielectric inclusions, as well as overall
surface defects, naturally leads to the time spread $t_d$ in
the cathode flare  formation in different microregions of the
cathode.

Dielectric inclusions not only initiate explosive electron
emission, but also sustain it ~\cite{Mesyats1993a}. The matter is
that the activity of each emitting center is accompanied by the
ion flow to the cathode (Fig.2). Dielectric inclusions in the
vicinity of the emitting center are charged by the ion current
resulting in the breakdown and the formation of new emitting
centers.
Another mechanism of  cathode plasma formation is associated with
the field desorption of the atoms absorbed on the cathode
surface~\cite{Litvinov1983}.
This occurs in the regions where the local electric field exceeds
$\sim 10^7 $V/cm. Collisional ionization of the desorbed atoms
 from the field-emission current results in the formation of plasma
 layer on the cathode surface.

Expansion of the conducting plasma of the cathode flares leads to
screening of the nearby regions on the cathode surface by a strong
electric field. The analysis given in
\cite{Mesyats1993c,Belomytzev1987,Belomytzev1980} shows that the
characteristic radius of the screened region is

\begin{equation}
 \label{eee1}
r_{scr}\approx 5\cdot10^2 \, U^{-3/4} \, i^{1/2}\, h_{ac},
\end{equation}
where $U$[V] is the applied voltage, $I$[A] is the cathode flare
current, and $h_{ac}$[cm] is the cathode-anode gap. For
$U=400$~kV, $i=10$~A, and $h_{ac}=1.5$~cm, the characteristic
radius  of the screened region, according to  \eqref{eee1}, is
$r_{scr}=0.15$cm.

Thus, the number of the explosive-emission centers   $N_e$
occurring simultaneously on the surface of the  cathode of radius
 $r_c=3$~cm can be estimated at $N_e=r_c^2/r_{scr}^2=60$. The
 number of concurrent explosive-emission centers, $N_e$, indicates the degree of
inhomogeneity of the beam's  transverse structure.

\section{Cumulation Mechanism}

As it has been stated in the previous section, explosive electron
emission begins with the formation and expansion of cathode
flares. Explosive electron emission is most intense from the
cathode protrusions, particularly from the cathode's inner edge
(Fig.3).
Coulomb repulsion causes the charged particles to rush to the
region free from the beam.
As a result, the accelerated motion of electrons toward the anode
comes alongside the radial motion to the cathode's  symmetry axis.
As a result,  the high-current beam density increases multifold on
the axis of the relativistic vacuum diode as compared to the
average current density in the cathode-anode gap. The reported cumulation mechanism are described for the ring-type cathode (the circular cathode with a hole, which coincides with the cathode exis). It should be noted that the cumulation mechanism doesn't depend on the hole position, i.e. whether  or not the hole axis coincides with the cathode axis.

\begin{figure}[ht]
\begin{center}
\resizebox{85mm}{!}{\includegraphics{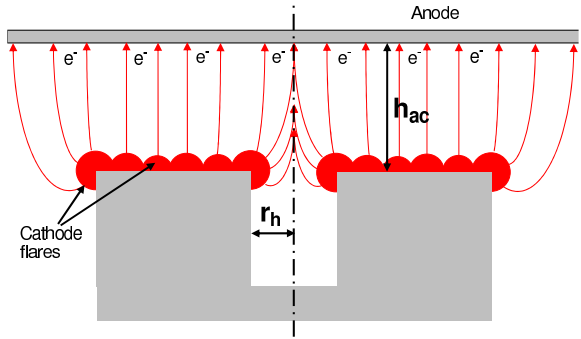}}\\
\end{center}
\caption{Cumulation caused by electrostatic repulsion} \label{Fig.3}
\end{figure}

Figure 4 shows the results of simulations: the dose absorbed by the anode. The assumed parameters of the
cathode were  as follows: cathode radius 3.0 cm, cathode-anode gap
2.0 cm, and the radius of the inner hole 0.8 cm.  The maximum
value of the accelerating voltage pulse was taken equal to  360 kV
and its duration -- to 330 ns. The simulated current density in
the region of the central spot on the anode at the moment
corresponding to the maximum accelerating voltage was as large as
1.0 kA/cm$^2$, being 5 times greater than the average current
density of the high-current diode. Thus, the simulation result
indicate the electron-beam cumulation on the axis of a
high-current diode with a ring-type cathode.

The undeniable advantage of this cumulation mechanism over a
conventional one based on the self-focusing of a high-current beam
by its own magnetic field  is  a very low energy spread of
particles in the region of the maximum current density  due to the
laminar flow of charged particles (Figure 5). In contrast, under the conditions of
self-focusing of the beam by its own magnetic field, the flow
current becomes appreciably  turbulent, and the charged particles
acquire a significant momentum spread~\cite{Poukey1974}. The
electron flow in this case is  like a compressed relativistic  gas
with electron temperature of the order of the voltage applied
across the diode~\cite{Rudakov1990}.

\begin{figure}[ht]
\begin{center}
\resizebox{85mm}{!}{\includegraphics{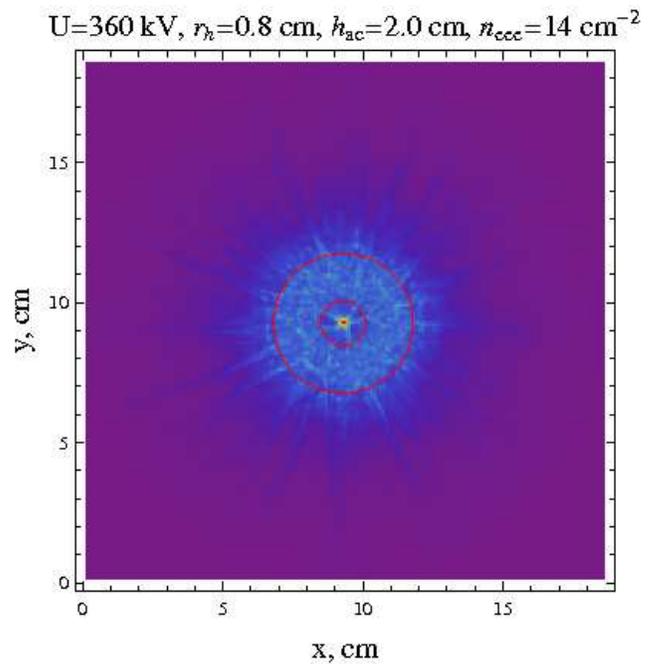}}\\
\end{center}
\caption{Absorbed dose (simulation results)} \label{Fig.4}
\end{figure}

\begin{figure}[ht]
\begin{center}
\resizebox{85mm}{!}{\includegraphics{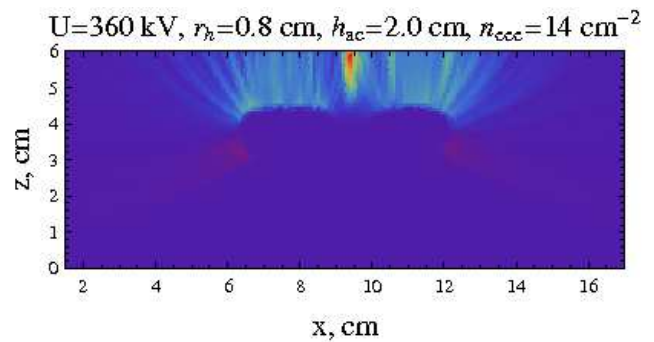}}\\
\end{center}
\caption{Electron flow (simulation results)} \label{Fig.5}
\end{figure}

\section{Experimental results}

To investigate the cathodes and obtain the information about
electron beam parameters, we used a nanosecond pulse-periodic electron
accelerator with a compact SF6-insulated high-voltage 
generator (HVG) as a power supply  providing pulsed voltage up to
400 kV in $\sim 30$ Ohm resistive load with half-height duration
of 130 ns and  rise time of 30 ns \cite{Bolshakov}. 

To obtain integrated full-sized imprints of the electron beam, we
used a radiochromic dosimetry film (technical specification TU 2379-026-13271746-2006)
\cite{Generalova} placed 3 mm behind the anode mesh made of
stainless steel (the geometrical transparency of the
mesh was 0.77); the cathode-anode gap was 2~cm. The dosimetry
film enabled us to
obtain information about the total absorbed dose due to the
passage of charged particles \cite{Pushkarev2005}. 

After the
exposure, the transmission scanning of the  dosimetry film using
the optical filter was made with EPSON Perfection V100 Photo
scanner. The distribution of the absorbed dose (and hence the
beam's energy density) over the beam cross section was derived
unambiguously from the scanned images and the dose-response
calibration curve.

\begin{figure}[ht]
\begin{center}
\resizebox{85mm}{!}{\includegraphics{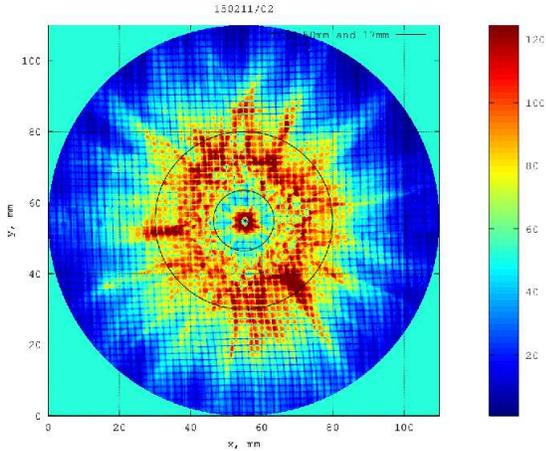}}\\
\end{center}
\caption{Electron beam imprint without foils} \label{Fig.6}
\end{figure}

Our first experiments showed that  the flow of charged particles
on the axis was so intense that it burned the film through (see
Fig.6). For this reason, we  placed 70 $\mu$m-thick  aluminium
foils in front of the dosimetry film to decrease the absorbed dose (see
Figs. 7--10).

\begin{figure}[ht]
\begin{center}
\resizebox{85mm}{!}{\includegraphics{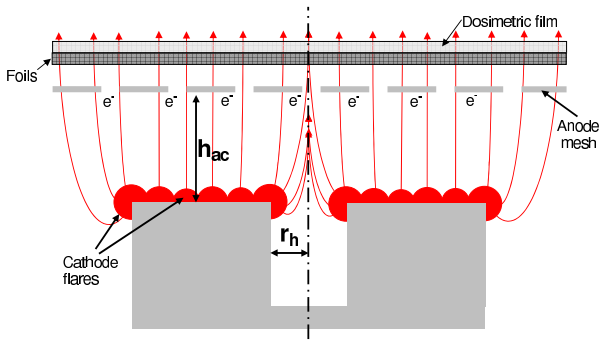}}\\
\end{center}
\caption{Experimental scheme} \label{Fig.7}
\end{figure}

\begin{figure}[ht]
\begin{center}
\resizebox{85mm}{!}{\includegraphics{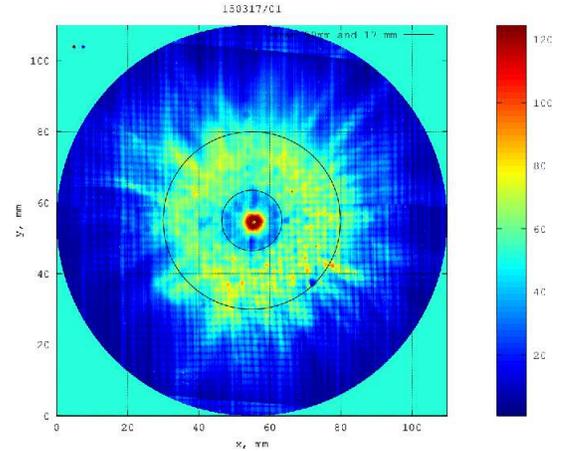}}\\
\end{center}
\caption{Electron beam imprint with one foil} \label{Fig.8}
\end{figure}

\begin{figure}[ht]
\begin{center}
\resizebox{85mm}{!}{\includegraphics{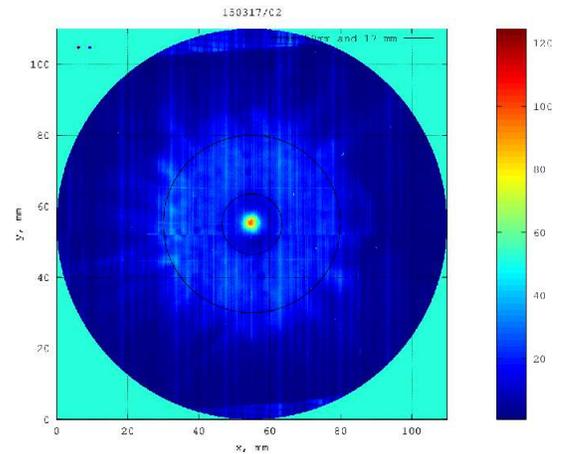}}\\
\end{center}
\caption{Electron beam imprint with two foils }
\label{Fig.9}
\end{figure}

\begin{figure}[ht]
\begin{center}
\resizebox{85mm}{!}{\includegraphics{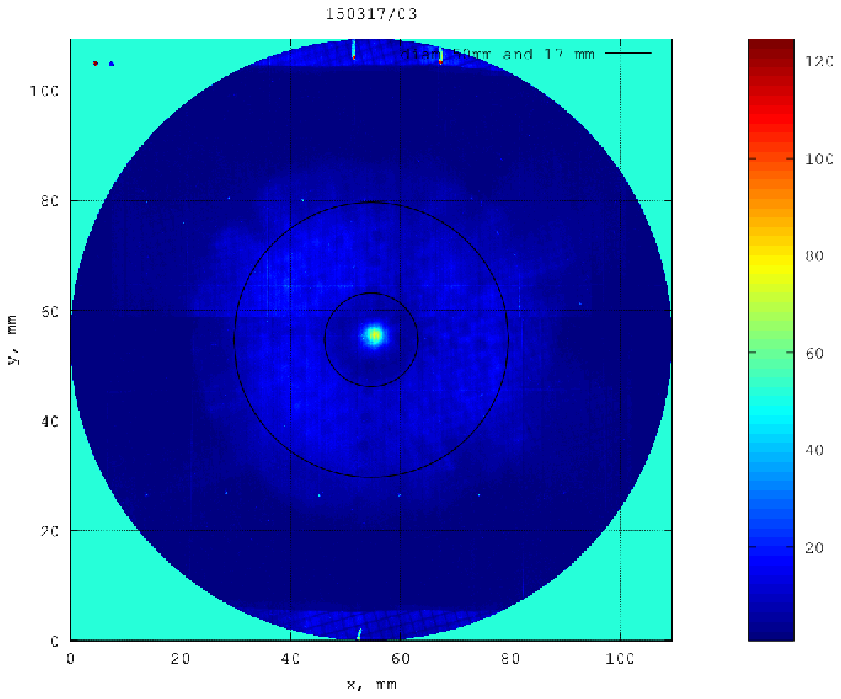}}\\
\end{center}
\caption{Electron beam imprint with three foils}
\label{Fig.10}
\end{figure}

 This enabled us  to cut off the  flows of both  the
weakly-relativistic electrons produced at the voltage pulse decay and the cathode plasma. The experiments conducted with
one, two, and three foils demonstrated that a sharp increase in
the absorbed dose remains in the center. This means that the
particle flow consists at the beam axis of high-energy electrons. In the
experiments with three foil layers cutting off all electrons whose
energy is less than 250 keV, the absorbed dose in the center was
almost four times as large as the average dose across the beam
cross section, showing a good agreement with the simulation
results.

Let us note here that both the simulation and the experiments were
performed at maximum accelerating voltage $\sim 400$\,kV and the
cathode-anode gap equal to 2 cm. The estimates show that with the
voltage increased to 2 MV and the cathode-anode gap decreased by a
factor of 5 it is possible to achieve the beam intensity of the
order of $1$TW/cm$^2$ required, for example, to study the extreme states of
matter and to do  research into  inertial confinement fusion.

\section{Conclusion}

In this paper we described the cumulation mechanism of a
high-current beam in a relativistic vacuum diode  with a ring-type cathode. The basis of this cumulation
mechanism is electrostatic repulsion of electrons from the
explosive-emission plasma  on the inner edge of the cathode.
The simulated values of  current density   and beam intensity
equal to $0.36$\,GW/cm$^2$ and 1 kA/cm$^2$, respectively,
qualitatively agree with the experimental data.

A very low particle energy spread in the region of maximum current
density that is due to laminar flow of charged particle is the
distinctive feature of the described cumulation mechanism over a
conventional one relying on focusing the high-current beam by its
own magnetic field.

\appendix
\section{Simulation of high-current beams}
In simulating of the electron beam dynamics under the conditions
of nonuniform explosive electron emission, it is necessary to
consider the expansion of the cathode plasma emitted from separate
explosive emission centers. Self-consistent simulation of particle
motion in self- and external electric and magnetic fields is
usually  performed using the particle-in-cell method
\cite{Roshal1979,Hockney1987,Birdsall1989} in a quasi-stationary
approximation \cite{Poukey1973,Golovin1989}. Quasi-stationary
approximation applies when the field parameters in high-current
diodes change slowly, and the displacement currents and induction
fields are neglected.

The simulation is performed  in the Cartesian, as well as in
cylindrical coordinates. The electric fields and particle motion
are computed in the Cartesian coordinate system, while the
magnetic fields - in cylindrical. Spatial dimensions of the cells
on the mesh for the field calculations are set equal, i.e.,
$\Delta X=\Delta Y=\Delta Z=\Delta R$. The system is assumed to be
axially symmetric, but the numerical simulation of particle motion
is performed in 3 dimensions, which is of principal necessity for
a proper consideration of the electron emission nonuniformity.

\begin{figure}[ht]
\begin{center}
\resizebox{85mm}{!}{\includegraphics{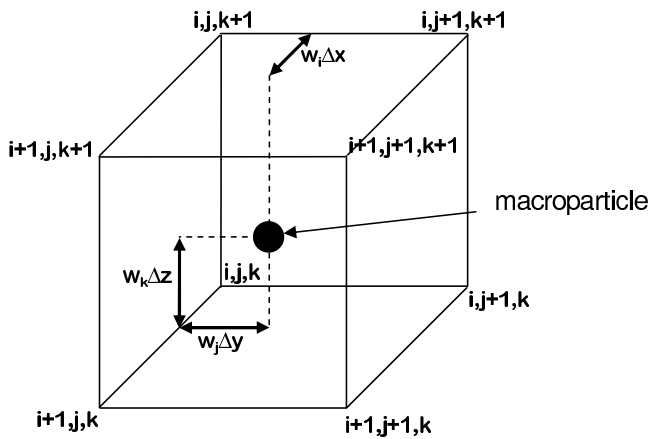}}\\
\end{center}
\caption{Particle-in-cell}
\label{Fig.10}
\end{figure}

\subsection{Particle-in-cell method}

The particle-in-cell method has been developed for simulating the
multiple phenomena in different fields of physics: vacuum
electronics \cite{Hartree1941,Hartree1950,Tien1956,Antonsen},
plasma
physics~\cite{Buneman1959,Lomax1960,Birdsall1961,Dawson1962,Langdon1973,Langdon1979,Hewitt1987,Sveshnikov1989,Birdsall1991,Verboncoeur2005},
hydrodynamics~\cite{Harlow1964}, magnetic hydrodynamics
\cite{Marder1975,Brunel1981}, astrophysics \cite{Hockney1967}, and
semiconductor physics~\cite{Hockney1974,Warriner1976}.

This method consists in the representation of real flows of
charged particles  (electrons, protons, and ions) as a set of
macroparticles, each containing a large number of real charge
carriers. Every macroparticle, normally located in a single cell,
has a certain attributed spatial distribution of mass and charge.
Depending on the charge and the location of the macroparticle,  a
certain contribution to the charge and current densities is
attributed to its nearest nodes on the spatial mesh using   the
weighting procedure.  Using a similar procedure, one can find the
forces acting on the macroparticle, knowing the magnitudes of the
fields in the nodes located in the close proximity to the
particle. Substituting the value for the force into the
finite-difference analogues  of relativistic equations, we may
find new locations and momenta of macroparticles. For a full
description of the system of fields and particles, these
procedures are completed with  charged particle injection  into
and removal from the computational region.

It is the numerically realized  injection of charged particles
emitted from the surface of the expanding cathode flares that
constitutes the novel and important feature of the developed code.

 A typical
program cycle based on the particle-in-cell method consists
of six operations: computation of coordinates and momenta of
particles, injection and removal of particles, computation of the
current and charge densities, and computation of the electric and
magnetic fields. Now let us proceed to a detailed description of
each procedure realized in our code.

\subsection{Electric fields}

The electric field strength in the quasi-stationary approximation
in the Coulomb gauge   is related to the scalar potential $\phi$
governed by the Poisson equation:
\begin{equation}
 \label{pd1}
\Delta\phi=-4\pi\rho,
\end{equation}
\begin{equation}
 \label{pd2}
\vec E=-\nabla\phi.
\end{equation}
The analysis given in  \cite{AG2014}  showed that the Jacobi
iterative method fits best to solve  the Poisson equation for
plasma dynamical problems. It is well known  \cite{Hockney1987}
that the iterative methods requiring that  the approximate value
of the potential at the first iteration step be  specified are
slowly converging, because the initial distribution usually
deviate appreciably from the exact solution of the
finite-difference analogue of the Poisson equation. In plasma
dynamical problems the situation is basically different, which
stems from a more appropriate selection of  the initial
approximation at the first iteration step: the grid potential
magnitudes obtained at the previous time step are taken as the
first approximation \cite{AG2014}. As a result, the entire
iteration process at each time step reduces to one-three
iterations,  requiring much less time than, say, the computation
of new coordinates and positions of the particles.

Thus, with the Jacobi iterative method the final-difference
analogue of the Poisson equation has the form:
\small\begin{equation}
\label{pd3}
\begin{split}
&
\frac{\phi_{i+1,j,k}^{n,s}-2\phi_{i,j,k}^{n,s+1}+\phi_{i-1,j,k}^{n,s}}{\Delta
X^2}+
\frac{\phi_{i,j+1,k}^{n,s}-2\phi_{i,j,k}^{n,s+1}+\phi_{i,j-1,k}^{n,s}}{\Delta Y^2}\\
&+\frac{\phi_{i,j,k+1}^{n,s}-2\phi_{i,j,k}^{n,s+1}+\phi_{i,j,k-1}^{n,s}}{\Delta Z^2}=-4\pi\rho_{i,j,k}^{n},\\
&\phi_{i,j,k}^{n,0}=\phi_{i,j,k}^{n-1},\\
\end{split}
\end{equation}
 where  $s$  is the iteration number and  $n$ is the time step number. The iteration in  \eqref{pd3} occurs until the residual
  $|\phi^{n,s}-\phi^{n,s-1}|$ becomes less than
$\epsilon|\phi^{n,s}|$. The value of  $|\phi^{n,s}|$, which is the
norm of the $\phi^{n,s}$ matrix is computed by formula
\begin{equation}
 \label{pd4}
|\phi^{n,s}|=\sqrt{\sum_{i,j,k}(\phi^{n,s}_{i,j,k})^2}.
\end{equation}
The parameter $\epsilon$ in fact determines the error of the
finite-difference Poisson equation solution.

To solve the Poisson equation, we need to complete it with the
boundary conditions. We used the Dirichlet boundary condition
implying the specified potentials on the cathode  ($\phi=U_c$) and
the anode ($\phi=U_a$). At the edge of the computational region in
the cathode-anode gap we took the logarithmical distribution of
the potential  \cite{Bugaev1979}
\begin{equation}
 \label{pd5}
\phi=U_c+\frac{(U_a-U_c)\ln(r/r_c)}{\ln(r_a/r_c)},
\end{equation}
that exactly describes the change in  $\phi$ in the gap between
two infinite cylinders. Obviously, the distribution of the
potential in  a high-current diode  will approach  \eqref{pd6} if
the boundary of the computational region stated here is located at
a considerable distance from the electron-emitting surface.
 We shall find the grid density $\rho_{i,j,k}^{n}$ with the linear
 weighting procedure attributing a certain contribution to
 $\rho_{i,j,k}^{n}$ coming from eight nodes nearest to the particle located at $(x_\alpha,y_\alpha,z_\alpha)$
\begin{equation}
\begin{split}
&\Delta\rho_{i,j,k}^{\alpha n}=q_\alpha(1-w^n_i)(1-w^n_j)(1-w^n_k)/\Delta V\\
&\Delta\rho_{i+1,j,k}^{\alpha n}=q_\alpha w^n_i(1-w^n_j)(1-w^n_k)/\Delta V\\
&\Delta\rho_{i,j+1,k}^{\alpha n}=q_\alpha (1-w^n_i)w^n_j(1-w^n_k)/\Delta V\\
&\Delta\rho_{i,j,k+1}^{\alpha n}=q_\alpha (1-w^n_i)(1-w^n_j)w^n_k/\Delta V\\
&\Delta\rho_{i+1,j+1,k}^{\alpha n}=q_\alpha w^n_iw^n_j(1-w^n_k)/\Delta V\\
&\Delta\rho_{i,j+1,k+1}^{\alpha n}=q_\alpha (1-w^n_i)w^n_jw^n_k/\Delta V\\
&\Delta\rho_{i+1,j,k+1}^{\alpha n}=q_\alpha w^n_i(1-w^n_j)w^n_k/\Delta V\\
&\Delta\rho_{i+1,j+1,k+1}^{\alpha n}=q_\alpha w^n_i w^n_j w^n_k/\Delta V.\\
\end{split}
\end{equation}
Here  $\Delta V=\Delta X\Delta Y\Delta Z$ и $\vec
w^n=\Big((x_\alpha^n-x_{i,j,k})/\Delta
X,(y_\alpha^n-y_{i,j,k})/\Delta Y, (z_\alpha^n-z_{i,j,k})/\Delta
Z\Big)$.

After we find the potential, the electric field is found
immediately from ~\cite{Volkov}
\begin{equation}
\label{pd6} 
\begin{split}
&\vec E^{n}_{i,j,k}=-\bigg(\frac{\phi^n_{i+1,j,k}-\phi^n_{i-1,j,k}}{2\Delta
X},\frac{\phi^n_{i,j+1,k}-\phi^n_{i,j-1,k}}{2\Delta Y},\\
&\frac{\phi^n_{i,j,k+1}-\phi^n_{i,j,k-1}}{2\Delta Z}\bigg).\\
\end{split}
\end{equation}

\subsection{Magnetic fields}

In considering motion of relativistic charged particles, it is
fundamentally important to take account of the self- and external
magnetic-field effect on the electron-beam dynamics in a
high-current diode. As we are concerned with axially symmetric
configurations, we shall proceed to cylindrical coordinates. By
virtue of axial symmetry, we shall assume that the magnetic field
$\vec H$ and the current density $\vec j$ are independent of the
azimuth angle $\theta$.

In the absence of the axial field $H_z^{ext}$,
 $H_\theta$ becomes the only magnetic field component and is
 related to the current density $j_z$ and the current $I$ running through the cathode by the Stokes theorem
\begin{equation}
 \label{pd8}
 H_\theta=\frac{4\pi(I(r,z)+2\pi\int_0^rj_z(r_1,z)r_1dr_1)}{2\pi cr}.
\end{equation}
Here the current $I(r,z)$ contains the contributions coming from
all electrons injected  into the points with coordinates  $>z$.

\subsection{Cathode plasma expansion }

The center of explosive emission is formed on the cathode when the
electric field strength exceeds a certain threshold value $E_{cr}$
that depends on the condition of the electrode surface. Then the
cathode flare begins to  expand  at a speed
$v_{eee}\sim2\cdot10^6$\,cm/s in many directions. Without
deliberate control over the surface microstructure, the emission
centers are chaotically located about the cathode surface. The
mean distance between them  is determined by the size of the
screened area $d_{scr}=2r_{scr}\sim3$~mm (see (\ref{eee1})),
knowing which we can easily estimate the characteristic density
$n_{eee}$ of the explosive-emission centers $n_{eee}=1/\pi
r_{scr}^2\sim14$~cm$^{-2}$.

For the purposes of high-current diodes simulation, we shall
assume that the emission regions are formed in the cathode nodes
where the electric field exceeds $E_{cr}$ with the probability
 $n_{eee}(\Delta X\Delta Y\Delta Z)^{2/3}$. When the electric
 field becomes greater than $E_{cr}$, the
 cathode flare expands at a constant speed $v_{eee}$ in every direction from the emission region.
 Each cathode flare is the source of electrons. The active cathode
 flare emits thermionic emission current which is many times as
 large as the current limited by the beam space charge, and so
 we can speak about the unlimited thermionic emission resulting in
practically zero field on the surface of the expanding cathode
plasma.  The assumption about the cathode screening enables us to
appreciably simplify the numerical computation of the
charged-particle kinetics, sparing ourselves the need to simulate
fast processes just in the emission region. Such simulation would
require a high space-time resolution  due to the smallness of the
Debye length and high values of plasma oscillation frequency
 \cite{AG2014}.

At each time step, we inject the  charged particles  into
plasma-occupied nodes that have in the vicinity at least one node
free from the conducting material.  The magnitude of the injected
charge $Q_{i,j,k}$ is found from the relation
\begin{equation}
\label{pd13}
\begin{split}
& Q_{i,j,k}=\Big(\frac{E_{xi+1,j,k}^n-E_{xi-1,j,k}^n}{2\Delta X}+
 \frac{E_{xi,j+1,k}^n-E_{xi,j-1,k}^n}{2\Delta Y}\\
&+\frac{E_{xi,j,k+1}^n-E_{xi,j,k-1}^n}{2\Delta Z}\Big)\frac{\Delta X\Delta Y\Delta Z}{4\pi}.\\
\end{split}
\end{equation}
Let us note that the charge is injected if  $Q_{i,j,k}<0$.

\subsection{Motion of charged particles}

Numerical integration of relativistic equations of motion is the
most time-consuming procedure of all, that is why it is paid
special attention to by the developers of the codes. In a nonrelativistic case, the most
widely used is the leapfrog scheme
\cite{Birdsall1989,Verboncoeur2005}
\begin{equation}
\label{pd14}
\begin{split}
 &\vec p_\alpha^{n+1/2}=\vec p_\alpha^{n-1/2}+\vec F_\alpha^{n}\Delta T,\\
 &\vec r_\alpha^{n+1}=\vec r_\alpha^n+\frac{\vec p_\alpha^{n+1/2}}{m_\alpha}\Delta T,\\
\end{split}
\end{equation}
Because the magnetic fields may be
neglected, the forces  $\vec F_\alpha^{n}=q_\alpha\vec E_\alpha^n$
contain only the electric field, their magnitudes being
unambiguously defined by the positions of particles and boundary
conditions.  In the relativistic case, the scheme  \eqref{pd14}
cannot be used directly, because $\vec F_\alpha^{n}$ includes the
Lorentz force  $q_\alpha \vec v_\alpha^n\times \vec H_\alpha^n$,
depending on the velocity  $\vec v_n$ determined at time
 $T_n=n\Delta T$. However, in the leapfrog scheme, the particle velocities are
 defined at half-integral times  $T_{n+1/2}=(n+1/2)\Delta T$. The natural solution allowing us
 to retain the simplicity of the leapfrog scheme in this
 situation is the application of Lagrange's interpolation  formula
 for computing $\vec v_\alpha^{n+1}$ from the three values of the
 velocity ($\vec
v_\alpha^{n-3/2}$, $\vec v_\alpha^{n-1/2}$, $\vec
v_\alpha^{n+1/2}$)

\begin{equation}
\label{pd15} \vec v_\alpha^{n+1}=\frac{3\vec
v_\alpha^{n-3/2}-10\vec v_\alpha^{n-1/2}+15\vec
v_\alpha^{n+1/2}}{8}.
\end{equation}

Thus, the complete integration scheme of the equations of motion
takes the form:
\begin{equation}
\label{pd16}
\begin{split}
 &\vec p_\alpha^{n+1/2}=\vec p_\alpha^{n-1/2}+q_\alpha(\vec E_\alpha^n+\vec v_\alpha^{n}\times \vec H_\alpha^n)\Delta T,\\
 &\vec v_\alpha^{n+1/2}=\frac{c\vec p_\alpha^{n+1/2}}{\sqrt{m_\alpha^2c^2+(\vec p_\alpha^{n+1/2})^2}}\\
 &\vec r_\alpha^{n+1}=\vec r_\alpha^n+\vec v_\alpha^{n+1/2}\Delta T,\\
&\vec v_\alpha^{n+1}=\frac{3\vec v_\alpha^{n-3/2}-10\vec v_\alpha^{n-1/2}+15\vec v_\alpha^{n+1/2}}{8}.\\
\end{split}
\end{equation}

The fields  $\vec E_\alpha^n$ and $\vec H_\alpha^n$ acting on the
particle are determined from the magnitudes of the mesh fields in
the eight adjacent nodes \cite{Verboncoeur2005}:
\begin{equation}
\begin{split}
& \vec E_\alpha^n=\vec E_{i,j,k}^n(1-w^n_i)(1-w^n_j)(1-w^n_k)\\
&+\vec E_{i+1,j,k}^nw^n_i(1-w^n_j)(1-w^n_k)\\
&+\vec E_{i,j+1,k}^n(1-w^n_i)w^n_j(1-w^n_k)\\
&+\vec E_{i,j,k+1}^n(1-w^n_i)(1-w^n_j)w^n_k\\
&+\vec E_{i+1,j+1,k}^nw^n_iw^n_j(1-w^n_k)\\
&+\vec E_{i,j+1,k+1}^n(1-w^n_i)w^n_jw^n_k\\
&+\vec E_{i+1,j,k+1}^nw^n_i(1-w^n_j)w^n_k\\
&+\vec E_{i+1,j+1,k+1}^n w^n_i w^n_j w^n_k.\\
\end{split}
\end{equation}


\end{document}